# Measuring university students’ satisfaction with traditional search engines and generative AI tools as information sources

## Brady D. Lund, Scott J. Warren, and Zoe Abbie Teel

### Abstract

This study examines university students’ levels of satisfaction with generative artificial intelligence (AI) tools and traditional search engines as academic information sources. An electronic survey was distributed to students at U.S. universities in late fall 2025, with 236 valid responses received. In addition to demographic information about respondents, frequency of use and levels of satisfaction with both generative AI and traditional search engines were measured. Principal components analysis identified distinct constructs of satisfaction for each information source, while k-means cluster analysis revealed two primary student groups: those highly satisfied with search engines but dissatisfied with AI, and those moderately to highly satisfied with both. Regression analysis showed that frequency of use strongly predicts satisfaction, with international and undergraduate students reporting significantly higher satisfaction with AI tools than domestic and graduate students. Students generally expressed higher levels of satisfaction with traditional search engines over generative AI tools. Those who did prefer AI tools appear to see them more as a complementary source of information rather than a replacement for other sources. These findings stress evolving patterns of student information seeking and use behavior and offer meaningful insights for evaluating and integrating both traditional and AI-driven information sources within higher education.

**Keywords:** information seeking; information sources; search engines; generative AI; university students; student satisfaction; academic libraries; information services; information searches; artificial intelligence; ChatGPT

All authors: University of North Texas, College of Information, Denton, Texas, USA.

**Abstract**

This study examines university students' levels of satisfaction with generative artificial intelligence (AI) tools and traditional search engines as academic information sources. An electronic survey was distributed to students at U.S. universities in late fall 2025, with 236 valid responses received. In addition to demographic information about respondents, frequency of use and levels of satisfaction with both generative AI and traditional search engines were measured. Principal components analysis identified distinct constructs of satisfaction for each information source, while k-means cluster analysis revealed two primary student groups: those highly satisfied with search engines but dissatisfied with AI, and those moderately to highly satisfied with both. Regression analysis showed that frequency of use strongly predicts satisfaction, with international and undergraduate students reporting significantly higher satisfaction with AI tools than domestic and graduate students. Students generally expressed higher levels of satisfaction with traditional search engines over generative AI tools. Those who did prefer AI tools appear to see them more as a complementary source of information rather than a replacement for other sources. These findings stress evolving patterns of student information seeking and use behavior and offer meaningful insights for evaluating and integrating both traditional and AI-driven information sources within higher education.

University students have found the ways that they search for, evaluate, and use information change dramatically with the emergence of generative artificial intelligence (AI) tools. While traditional search engines like Google had been dominant information sources for over two decades, the expansion of generative AI tools – including large language models like the Generative Pre-Trained Transformer and conversational platforms like ChatGPT – have introduced a new class of information sources that produce synthesized, human-like responses to queries. This development represents a major departure from information provided by search engines, which existed in the form of a curated list of web resources (Divekar et al., 2025). Given these notable shifts in information sources, it is critical to expand our understanding of how students perceive and use generative AI compared to traditional search engines, which types of information sources they generally prefer, and what factors may contribute to their adoption of these systems.

This paper addresses the gap in our understanding of generative AI as an information source through a comparative survey of U.S.-based university students. Through the collection of data about patterns of information use, satisfaction, and demographic variables, new insights are provided into the evolving relationship between AI-powered and traditional information sources in higher education. These findings have significant implications for educators, educational administrators, academic librarians, and technology designers when considering the interaction of their students with the various tools offered in their information environment.

## Literature Review

### Generative AI in Academic Contexts: Opportunities and Challenges

Generative artificial intelligence (GenAI) technologies have become an essential component of academic work in higher education. Students and faculty now use large language and text-to-image models to assist with writing, problem-solving, and design. These systems help both groups accelerate concept development, draft production, and iterative refinement, often improving efficiency and accessibility for learners who might otherwise face linguistic or disciplinary barriers (Kasneci et al., 2023; Mayer, 2021; Plass et al., 2020). GenAI may support students who struggle with academic language or conceptual articulation by scaffolding early stages of writing and idea generation.

However, these technological affordances also introduce challenges that influence the quality and credibility of academic work. GenAI systems frequently generate fabricated citations or subtle inaccuracies that require users to engage in careful verification and source checking (Epstein et al., 2023; Gasser & Mayer-Schönberger, 2024). Bias in training data can reinforce stereotypes if users fail to apply critical oversight during the creative process (Bianchi et al., 2023; Naik & Nushi, 2023). Intellectual property concerns remain unresolved, particularly regarding ownership of AI-generated content and the legality of copyrighted materials used during model training (Lemley & Casey, 2021; Sobel, 2020). These ethical, technical, and legal questions require scholars and students to act as both consumers and auditors of AI outputs.

Institutional policy and pedagogy have begun to respond to these developments. Instructors are revising assignments to emphasize transparency, process documentation, and independent reasoning (Kasneci et al., 2023). A recent report from Ithaka S+R (Baytas & Ruediger, 2025) found that many universities are moving toward "AI-integrated learning design," where students must document how generative systems were used and justify their outputs. People now view AI literacy, which includes prompt engineering, bias detection, and proper citation practices, as an essential skill alongside information literacy. In this sense, GenAI functions less as a replacement for human expertise and more as a cognitive partner that supports iterative thinking. Effective use depends on intentionality, awareness of limitations, and the user's capacity to critically evaluate and revise what AI produces.

**Student Adoption Patterns and Usage Behaviors**

Student adoption of GenAI has grown rapidly, though not uniformly across academic populations or disciplines. Surveys of university students indicate that adoption levels range from moderate experimentation to near ubiquity. For example, a global survey of 40 universities across six regions found that while 84% of students had tried GenAI tools, fewer than half used them regularly for academic writing (Jin et al., 2025). Similarly, a national survey of 1,041 U.K. undergraduates reported that 92% had used AI tools for coursework in 2025, up from 66% the year before (Freeman, 2025). Students primarily used GenAI for paraphrasing, brainstorming, and idea development but continued to rely on search engines and library databases for credible sources and verification (Sousa & Cardoso, 2025).

Patterns of use and satisfaction vary by demographic characteristics and context. A Hong Kong study examining undergraduate and postgraduate students found that learners valued GenAI for its ability to reduce writer's block and provide structural feedback but expressed concern about overreliance and factual reliability (Chan & Hu, 2023). The Ithaka S+R study of 19 universities yielded analogous results, revealing that many students utilized AI writing tools while expressing concerns regarding complete text generation or the substitution of human authorship (Baytas & Ruediger, 2025). Across contexts, students generally treat GenAI as a cognitive tool that complements, not supplants, traditional information sources.

Satisfaction with GenAI also appears to rise with use frequency, indicating that comfort and familiarity improve perceived utility for specific tasks. International students frequently report higher satisfaction than domestic peers because GenAI supports translation, academic writing conventions, and content adaptation (Chan & Hu, 2023; Sousa & Cardoso, 2025). These findings align with sociocultural and linguistic capital theories suggesting that GenAI serves as a mediator of access to academic discourse (Bourdieu, 1991; Lantolf, 2006). Students continue to trust search engines more for accuracy and source reliability but increasingly value GenAI for efficiency and linguistic support. The coexistence of both tools points toward a hybrid information ecosystem rather than a competitive one.

Pedagogical framing plays a significant role in how students use GenAI. Studies report that when instructors explicitly address verification, citation, and ethical use, students adopt more selective and critical strategies (Kasneci et al., 2023; Sousa & Cardoso, 2025). Without such guidance, use often drifts toward surface-level outputs, summaries, or rephrasing, without deeper reflection. Across the emerging body of research, three factors repeatedly shape learning outcomes: frequency of use, user type, and degree of instructional structure. These factors determine whether GenAI enhances or undermines learning quality.

**Satisfaction as a Construct in Technology Adoption and Use**

Satisfaction remains central in understanding post-adoption behavior. The Technology Acceptance Model (Davis, 1989) identifies perceived usefulness and ease of use as precursors to initial adoption. Information Systems Continuance Theory (Bhattacherjee, 2001) extends this by framing satisfaction as a key driver of sustained engagement. Expectation-Confirmation Theory (Oliver, 1980) provides further depth: satisfaction occurs when user experience aligns with or exceeds expectations. In academic contexts, satisfaction represents a post-adoption evaluation of how well GenAI tools support academic tasks such as reading, writing, and idea generation. It differs from trust or usefulness because it captures both affective judgment and behavioral intent, the degree to which students feel the tool fulfills their needs and are willing to continue using it.

Within Uses and Gratifications Theory (Katz et al., 1973), satisfaction reflects how effectively technologies deliver the gratifications users seek. For example, search engines satisfy needs for breadth, credibility, and transparency, whereas GenAI satisfies needs for efficiency, synthesis, and interactivity. As students gain experience, satisfaction may reinforce habitual use (Ouellette & Wood, 1998). However, research suggests the relationship is not linear. Overreliance without critical engagement can lower satisfaction as users encounter factual errors or ethical ambiguities (Sousa & Cardoso, 2025). Studies in education and information systems both indicate that satisfaction mediates the relationship between perceived usefulness and continued use, with satisfied users more likely to persist in their technology adoption (Islam & Azad, 2015). In short, satisfaction functions as both a reflection of experience and a predictor of sustained engagement, shaping how GenAI becomes normalized within students' information practices.

**Demographic and Behavioral Factors in Technology Preferences**

Demographic characteristics shape adoption and satisfaction in ways that extend beyond access or familiarity. Gender, age, academic level, and international status all influence attitudes toward technology use (Venkatesh & Morris, 2000). Younger or less experienced students often show higher openness to experimentation but lower confidence in verifying accuracy. Graduate and professional students, by contrast, tend to adopt selectively for specific analytical or design purposes (Bewersdorff et al., 2025). International student status is also relevant. For learners working in non-native language settings, GenAI can function as a linguistic and cultural bridge, supporting their idiomatic phrasing, disciplinary conventions, and conceptual clarity (Chan & Hu, 2023; Sousa & Cardoso, 2025). This function parallels theories of epistemic access and disciplinary participation, where tools mediate learners' integration into academic discourse (Gee, 2012; Schraw & Olafson, 2008).

Behavioral variables also interact dynamically with satisfaction. For example, frequency of use can strengthen confidence and competence, enhancing satisfaction, but uncritical dependence may diminish it over time. This cyclical relationship suggests that satisfaction and frequency mutually reinforce one another up to an optimal threshold (Ouellette & Wood, 1998). Students who combine GenAI with search engines, using each for its strengths, typically report higher satisfaction than those who rely exclusively on one platform (Jin et al., 2025; Sousa & Cardoso, 2025). The interplay of demographic, behavioral, and contextual factors highlights the complexity of GenAI adoption and underscores the need for integrated analyses that move beyond single-variable models.

## Research Problem and Questions

The emergence of generative AI tools has begun to reshape how university students search for, evaluate, and use information. While traditional search engines like Google supply information in a way that facilitates browsing and evaluating multiple resources and perspectives on an issue, generative AI offers a synthesized, human-like response to any query that represents a fundamentally different mode of human interaction with information. Researchers have noted that a large proportion of university students are eager adopters and advocates of AI use (Bewersdorff et al., 2025; Obenza et al., 2024). Despite this growing adoption of AI tools in higher education, little is known about how students perceive and compare these tools to traditional search engines, what factors influence their satisfaction, and how demographic and behavioral characteristics shape their use. Without this understanding, educators, librarians, and technology developers may struggle to design effective information and AI literacy programs and support systems that reflect students' evolving information-seeking behaviors.

To address this existing gap in the literature, this study investigates the following research questions:

- What are university students' levels of satisfaction with the information supplied by traditional search engines and generative AI tools?
- What demographic and behavioral characteristics predict higher or lower satisfaction with generative AI tools and traditional search engines as information sources?
- Can distinct groups of students be identified based on their patterns of satisfaction and use of generative AI tools and search engines as information sources?

# Methods

This study utilized an electronic survey developed using Qualtrics. The survey was distributed using a convenience sampling approach to students enrolled at U.S.-based universities during the period of August-September 2025. Eight questions were used for the analysis in this study. The first five are demographic questions asking about students' gender identity, age, academic major, academic standing, and student status (domestic or international). This was followed by questions that asked participants to estimate how many times they had used traditional search engines and generative AI tools in order to find information in the last month. Finally, two multi-part Likert scale questions asked respondents to indicate how satisfied they were with the ability of traditional search engines and generative AI tools to provide specific types of information: job-related information, coursework-related information, information for studying topics, weather-related information, and news information.

Following data collection, the response data was transferred from Qualtrics to SPSS software for further processing and analysis. Among the types of analyses performed with the data were:

1. Principal Components Analysis to first identify the underlying structure among the five information satisfaction questions and create a singular construct that measures satisfaction across information types for traditional search engines and artificial intelligence.
2. K-means Cluster Analysis to determine if certain groups of respondents emerge based on their satisfaction with information provided by search engines and AI tools, and what their demographic compositions look like.
3. Regression Analysis to look more directly at what demographic attributes of respondents may predict their satisfaction with these information sources across all types of information needs.

This series of analyses provides deeper insights into potential factors contributing to the growth in satisfaction with traditional search engine information and information generated by AI tools. Together, they provide value context to our understanding of how users' backgrounds, patterns of behavior, and perceived satisfactions with technological tools are linked.

## Results

236 responses were received for the survey. Three-fourths of these respondents were from the natural and computing sciences and graduate-level students. The gender distribution was roughly equal (48% female, 43% male, 9% other) as was student domestic/international status (43% domestic, 57% international). Respondents also skewed younger, with 63% under age 25, 26% aged 26–30, and 11% aged 31 or older, though this distribution is unsurprising given that all respondents were university students.

### Information Source Satisfaction Constructs

Two constructs were identified based on responses to two sets of five questions (questions 7a-7e and 8a-8e in the Appendix). A Principle Components Analysis (PCA) revealed a single component with an eigenvalue of 2.84, explaining 52.66% of the total variance. All five variables loaded positively on this component, with loadings ranging from .622 to .768 (Table 1). Based on the Kaiser criterion (eigenvalue > 1) and visual inspection of the scree plot, only one component was retained, representing "satisfaction with traditional search engines as an information source."

**Table 1. Factor Loadings for Search Engine Satisfaction**

| Construct | Loading |
|---|---|
| Job-Related Information | .768 |
| Coursework-Related Information | .720 |
| Information for Studying Topics | .768 |
| Weather-Related Information | .622 |
| News Information | .741 |

PCA revealed a single component with an eigenvalue of 3.41, explaining 68.27% of the total variance. All five variables loaded positively on this component, with loadings ranging from .798 to .855 (Table 2). Based on the Kaiser criterion (eigenvalue > 1) and visual inspection of the scree plot, only one component was retained, representing "satisfaction with generative artificial intelligence as an information source."

**Table 2. Factor Loadings for Artificial Intelligence Satisfaction**

| Construct | Loading |
|---|---|
| Job-Related Information | .855 |
| Coursework-Related Information | .805 |
| Information for Studying Topics | .839 |
| Weather-Related Information | .798 |
| News Information | .832 |

## Comparison of Satisfaction with Search Engine and AI as Information Source

Overall satisfaction with traditional search engines and generative AI as information sources is measured on a scale of 0-4, where:

- 0 represents "I am never satisfied with this tool"
- 1 represents "I am rarely satisfied with this tool"
- 2 represents "I am sometimes satisfied with this tool"
- 3 represents "I am often satisfied with this tool" and
- 4 represents "I am always satisfied with this tool.

Table 3 displays the mean, standard deviation, median, range, and distribution statistics for satisfaction ratings for search engines and artificial intelligence tools. In general, satisfaction with search engines is higher than artificial intelligence, with the mean and median ratings being similar for search engines (3.12 and 3.25) and artificial intelligence (2.11 and 2.17) and the former equating to "I am often satisfied with this tool" while the later equates to "I am sometimes satisfied with this tool," a rather notable distinction.

**Table 3. Descriptive Statistics for Satisfaction with Search Engines and AI as Information Source**

| Statistic | Search Engine | Artificial Intelligence |
|---|---|---|
| Mean | 3.12 | 2.11 |
| Standard Deviation | 0.80 | 1.00 |
| Median | 3.25 | 2.17 |
| Lower Bound | .13 | .68 |
| Upper Bound | 4.00 | 4.00 |
| Skewness | -1.11 | -0.075 |
| Kurtosis | 1.13 | -1.16 |

Correlation between satisfaction with search engine and satisfaction with AI is -.22 ($p < .01$), suggesting that those who are more satisfied with search engines are less satisfied with artificial intelligence and vice versa. Correlation between the number of times per month one uses a search engine and their satisfaction with search engines is moderately strong and positive at .33 ($p < .01$), while the relationship with satisfaction with AI is statistically insignificant and negative at -.10 ($p = .13$). Correlation between the number of times per month one uses AI and their satisfaction with AI is strong and positive at .58 ($p < .01$), while the relationship with satisfaction with search engines is statistically insignificant at .03 ($p = .61$).

### *Categorization of Students Based on Information Source Satisfaction*

To better understand the patterns of satisfaction between traditional search engines and generative AI tools, a k-means cluster analysis was performed. This method grouped students into clusters based on their reported satisfaction levels with each information source. Based on this clustering approach, two distinct groups of students emerged:

Cluster 1: Satisfaction of AI mean = 1.08; Satisfaction of SE mean = 3.48 (41% cases)

Cluster 2: Satisfaction of AI mean = 2.82; Satisfaction of SE mean = 2.88 (59% cases)

Cluster 1 represents a group of students who remain heavily reliant on traditional search engines and report being generally dissatisfied with generative AI as an information source. Cluster 2 captures a more balanced group of students who are open to both technologies, expressing a moderate-to-high level of satisfaction with each. Notably, there is no third cluster of students who are heavily reliant on AI and dissatisfied with search engines. Those who have satisfaction with AI still generally have positive perceptions of traditional search engines as well.

There are some key demographic distinctions between those who fall in cluster 1 and cluster 2. Attributes of cluster 1 (low AI satisfaction, high SE satisfaction):

- Includes 60.4% of domestic students but only 17.6% of international students.
- Includes 64.7% of graduate students but only 23.2% of undergraduate students.
- Includes 64.4% of women and 51.8% of men.

Attributes of cluster 2 (high AI satisfaction and high SE satisfaction):

- Includes 39.6% of domestic students and 82.4% of international students.
- Includes 35.3% of graduate students and 76.8% of undergraduate students.
- Includes 35.6% of women and 48.2% of men.

## Predicting Satisfaction with Search Engines as an Information Source

Regression analyses were performed for the satisfaction levels with search engines and generative AI tools as information sources based on predictor variables including user demographics and times which the technology tools were used in the prior month. The regression statistics are presented in Tables 4 (search engine) and 5 (AI).

For satisfaction with search engines as an information source, a significant model was found with values of R2 = .141, F = 6.11, $p < .01$. Only significant factor is the number of times in which one has used search engines in the past month. For each time that one used a search engine, the satisfaction with this tool increased by .03, meaning about 30 uses per month would distinguish some who is sometimes satisfied on search engines and someone who is very frequently satisfied with search engines, for instance.

**Table 4. Regression Statistics for Satisfaction with Search Engines as an Info Source**

| Variable | Unstandardized Beta | Standardized Beta | Significance |
| --- | --- | --- | --- |
| Constant | 2.53 | | <.01 |
| Academic Major (Computing) | .08 | .05 | .59 |
| Academic Standing (Graduate) | .09 | .04 | .58 |
| Age (26+) | -.03 | -.02 | .80 |
| Gender (Male) | -.00 | -.00 | .99 |
| Student Status (International) | -.26 | -.16 | .07 |
| Times Used SE Past Month | .03 | .34 | <.01 |

**Predicting Satisfaction with Artificial Intelligence as an Information Source**

For satisfaction with artificial intelligence tools as an information source, a significant model was found with values of R2 = .453, F = 30.63, p < .01. Several factors significantly effect satisfaction with AI as an information source. Graduate students have a lower satisfaction by a factor of -.53. International students have a higher level of satisfaction of .82. For each use of AI in the past month, satisfaction increases by a factor of .04. For instance, a domestic graduate student who never uses AI would have a projected satisfaction of 1.13 (1.66-.53+0.00+0.00), while an international undergraduate student who uses AI 20 times per month would have a projected satisfaction of 3.28 (1.66-0.00+.82+.80), a substantial difference of 2.15, or the difference between "rarely satisfied" and "often satisfied."

**Table 4. Regression Statistics for Satisfaction with Generative AI as an Info Source**

| Variable | Unstandardized Beta | Standardized Beta | Significance |
| --- | --- | --- | --- |
| Constant | 1.66 | | <.01 |
| Academic Major (Computing) | -.11 | -.05 | .42 |
| Academic Standing (Graduate) | -.53 | -.17 | <.01 |
| Age (26+) | .16 | .08 | .18 |
| Gender (Male) | -.04 | -.02 | .67 |
| Student Status (International) | .82 | .40 | <.01 |
| Times Used AI Past Month | .04 | .39 | <.01 |

## Discussion

### Satisfaction with Using AI

This study shows that university students tend to have greater satisfaction with using traditional search engines for finding information than with generative AI tools or large language models. However, the surface-level numbers do not tell the whole story, as there is nuance in the factors that predict greater satisfaction with AI as an information source. While demographics such as age, gender, and academic major do not appear to play a major role, academic standing and international status do.

### Undergraduate and International Student Perspectives

Undergraduates and international students generally express higher satisfaction with AI than their peers in this study. One possible explanation is that these groups may prefer information that is presented clearly and concisely, without the need to sort through multiple sources or interpret dense academic language. In contrast, domestic and graduate students may be more accustomed to navigating academic databases or conducting in-depth searches, or may be less comfortable with using AI tools to support their work.

International and undergraduate students may also value the accessibility and conversational nature of AI tools, which can make complex topics feel more approachable and less intimidating (Holmes et al., 2023). As mentioned previously, Chan & Hu (2023) found that "non-native English-speaking students" benefit from GenAI for translation and rewriting tasks. This aligns with the findings of this study. In addition, Deng, Liu, & Zhai (2025) reported that non-native English speakers, and those classified as "international," used and perceived more benefit from GenAI overall. This does not imply a lack of ability; rather, it may reflect that these students are at a different stage in their academic development, where intuitive, interactive tools help bridge gaps in confidence and research experience. These patterns align with existing research on college students' information-seeking strategies, which highlights a tendency to prioritize convenience and efficiency over depth or quality when conducting online searches for academic purposes (Abogdera, 2022).

For international students in particular, AI can be especially appealing because it offers clear explanations in natural, conversational language and has the ability to rephrase, simplify, or translate complex ideas—features that traditional databases typically do not provide (Wang et al., 2023). Similarly, many undergraduate students juggle multiple responsibilities such as coursework, employment, and adapting to college life (Mills, 2020). For them, AI's concise and synthesized responses can reduce cognitive load and save valuable time compared to sorting through numerous dense academic sources.

**Graduate and Domestic Student Perspectives**

On the other hand, graduate and domestic students might place greater value on traditional search engines and academic databases because they are trained to prioritize source credibility, methodological rigor, and peer-reviewed material. This emphasis on verifiable and citable sources aligns closely with the expectations of advanced academic work, where accuracy and traceability are essential. While graduate students increasingly recognize AI as a valuable tool for enhancing efficiency and supporting the quality of research (Xiang et al., 2024), findings from this study indicate that they still tend to prefer traditional search engines as their primary information source.

Graduate students, for instance, often conduct literature reviews, synthesize empirical findings, or engage in discipline-specific research that requires direct access to original studies. These tasks demand precision and source transparency, which traditional databases such as Google Scholar and library databases are specifically designed to provide (Zhao et al., 2023; Liu, 2023). Similarly, domestic students who have spent more time in U.S. academic settings may be more familiar with institutional library systems, citation practices, and evaluating source reliability; as Burton and Chadwick (2000) note, approximately 60% of domestic students (in their study) receive library research training in high school, giving them an early advantage in navigating these various resources.

## Use & Satisfaction with AI as an Information Source

A clear positive relationship between use frequency and satisfaction emerged for both information sources, particularly for AI. This feedback loop suggests that familiarity breeds confidence: students who experiment more with AI tools tend to perceive them as more useful and reliable. Conversely, those who rarely use AI may lack the experience needed to interpret or verify its outputs effectively, reinforcing skepticism (Mulford, 2025). Encouraging guided exploration and reflective use of AI in academic settings could therefore help students develop a more balanced understanding of its strengths and limitations. Recent surveys also show that while most university students report experimenting with AI tools, satisfaction and perceived trustworthiness are strongly correlated with personal experience and guided exposure (Mulford, 2025).

## Relationship Between AI and Traditional Search Tools

Interestingly, the relationship between satisfaction with AI and traditional search engines is not entirely reciprocal. Students who hold particularly favorable views of traditional search engines tend to express lower satisfaction with AI tools; however, those who are highly satisfied with AI-generated information generally maintain positive attitudes toward traditional search engines as well. This suggests that students who have embraced AI are not necessarily abandoning traditional information sources but rather integrating AI as a complementary component within their broader information-seeking habits. In other words, for many students, AI appears to serve as an additional resource that enhances—rather than replaces—their existing research and information practices.

## Educational Implications

These findings carry meaningful implications for educators, librarians, and institutional leaders. Students' varying levels of satisfaction with AI and search engines reveal differences in digital confidence, information literacy, and academic readiness. Rather than discouraging AI use, institutions should leverage this insight to design instruction that cultivates both AI literacy and critical information evaluation skills. For instance, integrating AI-supported search tasks into coursework can help students practice verifying AI outputs against peer-reviewed sources—transforming AI from a "shortcut tool" into a structured learning aid (Yin, Li, & Yu, 2023).

Supporting international and early-career students is particularly important. For these groups, AI can serve as a valuable accessibility and scaffolding mechanism, but guidance is needed to ensure responsible use and awareness of AI's biases and limitations. By embedding such instruction into first-year seminars or writing courses, universities can promote equitable digital competence across diverse student populations (Wang et al., 2023).

## Limitations and Future Research

While this study provides valuable insights into how university students perceive and evaluate generative AI and traditional search engines as information sources, several limitations should be acknowledged. First, the sample relied on a convenience-based, self-selected group of students from U.S. universities, which may limit the generalizability of the results. Respondents were disproportionately drawn from the natural and computing sciences and included a majority of graduate-level participants, potentially skewing satisfaction and adoption patterns toward students who are more technologically confident or academically experienced. Future studies may seek a more balanced representation across disciplines and academic levels, particularly by including students in the humanities, education, and social sciences, who may engage with AI tools in different ways.

Furthermore, this study utilized self-reported measures of satisfaction and frequency of use, which may be influenced by recall bias or social desirability effects. Students who perceive AI positively may overestimate their use or satisfaction, while those who view it skeptically may underreport use. Incorporating behavioral data, such as system logs, real-time task tracking, or experimental designs, could provide more accurate insight into how students actually interact with AI and search engines when performing academic tasks.

Finally, this study examined satisfaction at a general level across various information needs (e.g., coursework, job-related, and news information). Future research could differentiate among these contexts to determine whether satisfaction varies by task type or disciplinary requirement. Qualitative approaches such as focus groups or think-aloud protocols would further illuminate how students conceptualize “trust,” “usefulness,” and “credibility” when comparing AI outputs with traditional search results. Expanding the research to include international comparative samples would also help identify cultural or linguistic factors that mediate AI adoption and satisfaction.

## Conclusion

While generative AI tools are clearly becoming an increasingly ubiquitous part of the academic information landscape, the findings of this study suggest that university students as a whole remain more satisfied with traditional search engines as an information source. Students appear to value the reliability and perceived credibility of search engines – providing a diversity of resources for any query – even as they actively use AI tools for certain tasks that benefit from synthesized or conversational responses. These perceptions are not uniform among all students, as international and undergraduate students report significantly higher satisfaction with AI as an information source than domestic and graduate students. Yet, even when students rate AI tools highly as an information source, they tend to rate search engines highly as well, suggesting that AI tools are not simply replacing search engines but rather are being integrated as complementary resources by students who find both types of resources useful.

University administrators, educators, academic librarians, and developers may interpret these results to highlight a need for continued information literacy initiatives that incorporate AI literacy as one component of the modern information ecosystem. Teaching students how to critically evaluate AI-generated content while also emphasizes traditional strategies for effective searches and information source evaluation is critical to help them succeed in the education and future career activities. Through better understanding the demographic and behavioral factors that might shape student information source preferences, higher education institutions can further support a broad range of information seeking behaviors that create informed engagement in the modern technological landscape.

## References

Abogdera, A. (2022). *Exploring Information-Seeking Strategies College Students Use to Improve the Relevance of Retrieval from Online Information Retrieval Systems* (Doctoral dissertation, Colorado Technical University).

Baytas, C., & Ruediger, D. (2025). *Making AI generative for higher education: Adoption and challenges among instructors and researchers* (SR Report). Ithaka S+R. https://doi.org/10.18665/sr.322677

Bewersdorff, A., Hornberger, M., Nerdel, C., & Schiff, D. S. (2025). AI advocates and cautious critics: How AI attitudes, AI interest, use of AI, and AI literacy build university students' AI self-efficacy. *Computers and Education: Artificial Intelligence, 8*, 100340. https://doi.org/10.1016/j.caeai.2024.100340

Bhattacherjee, A. (2001). Understanding information systems continuance: An expectation-confirmation model. *MIS Quarterly, 25*(3), 351–370. https://doi.org/10.2307/3250921

Bianchi, F., Kalluri, P., Durmus, E., & Caliskan, A. (2023). Easily accessible text-to-image generation amplifies demographic stereotypes on a large scale. *Proceedings of FAccT*, 1493–1504. https://doi.org/10.1145/3593013.3594095

Bourdieu, P. (1991). *Language and Symbolic Power*. Harvard University Press.

Chan, C. K. Y., & Hu, W. (2023). Students' voices on generative AI: Perceptions, benefits, and challenges in higher education. *International Journal of Educational Technology in Higher Education, 20*, Article 43. https://doi.org/10.1186/s41239-023-00411-8

Davis, F. D. (1989). Perceived usefulness, perceived ease of use, and user acceptance of information technology. *MIS Quarterly, 13*(3), 319–340. https://doi.org/10.2307/249008

Deng, N., Liu, E. J., & Zhai, X. (2025, July). Understanding university students' use of generative AI: The roles of demographics and personality traits. In *International Conference on Artificial Intelligence in Education* (pp. 281-293). Cham: Springer Nature Switzerland.

Divekar, R. R., Gonzalez, L., Guerra, S., Boos, N., & Divekar, R. (2025). Can generative AI replace search engines for learning? Understanding student preferences, user and perceived proficiency in using AI. TechTrends, early view. https://doi.org/10.1007/s11528-025-01095-9

Epstein, Z., Hertzmann, A., Akten, M., & Teevan, J. (2023). Art and the science of generative AI. *Science, 380*(6650), 1110–1111. https://doi.org/10.1126/science.adh4451

Freeman, J. (2025). *Student generative AI survey 2025* (HEPI Policy Note 61). Higher Education Policy Institute. https://www.hepi.ac.uk/2025/02/26/student-generative-ai-survey-2025/

Gasser, U., & Mayer-Schönberger, V. (2024). *Guardrails: guiding human decisions in the age of AI.* Princeton University Press.

Gee, J. P. (2012). *Social Linguistics and Literacies: Ideology in Discourses* (4th ed.). Routledge.

Holmes, W., Bialik, M., & Fadel, C. (2023). Artificial intelligence in education. Data Ethics. Building Trust: How Digital Technologies can serve humanity, 621–653. *Globethics Publications. https://doi. org/10.58863/20.500*, *12424*, 4276068.

Islam, A. K. M. N., & Azad, N. (2015). Satisfaction and continuance with a learning management system: Comparing perceptions of educators and students. *International Journal of Information and Learning Technology, 32*(2), 109–123. https://doi.org/10.1108/IJILT-09-2014-0020

Jin, Y., Yan, L., Echeverria, V., Gašević, D., & Martinez-Maldonado, R. (2025). Generative AI in higher education: A global perspective of institutional adoption policies and guidelines. *Computers and Education: Artificial Intelligence, 8*, Article 100362. https://doi.org/10.1016/j.caeai.2024.100348

Kasneci, E., Seßler, K., Küchemann, S., & Kasneci, G. (2023). ChatGPT for good? Opportunities and challenges for education. *Learning and Individual Differences, 103*, 102274. https://doi.org/10.1016/j.lindif.2023.102274

Katz, E., Blumler, J. G., & Gurevitch, M. (1973). Uses and gratifications research. *The Public Opinion Quarterly, 37*(4), 509–523. https://doi.org/10.1086/268109

Lantolf, J. P. (2006). Sociocultural theory and L2: State of the art. *Studies in Second Language Acquisition, 28*(1), 67–109. https://doi.org/10.1017/S0272263106060037

Lemley, M. A., & Casey, B. (2021). Fair learning. *Texas Law Review, 99*(4), 743–805. https://texaslawreview.org/fair-learning/

Liu, Q. (2023). Information literacy and recent graduates: Motivation, self-efficacy, and perception of credit-based information literacy courses. *The Journal of Academic Librarianship*, 49(3), 102682.

Mayer, R. E. (2021). *Multimedia Learning* (3rd ed.). Cambridge University Press.

Mills, L. (2020). Understanding the experiences of college students who work full-time: Juggling competing responsibilities and defining academic success. The Journal of Continuing Higher Education, 68(3), 181-189.

Mulford, D. (2025). *AI in Higher Education: A Meta Summary of Recent Surveys of Students and Faculty*. Campbell Academic Technology Services. https://sites.campbell.edu/academictechnology/2025/03/06/ai-in-higher-education-a-summary-of-recent-surveys-of-students-and-faculty/?utm_source=chatgpt.com

Naik, R., & Nushi, B. (2023). Social biases through the text-to-image generation lens. *AIES 2023*, 786–808. https://doi.org/10.1145/3600211.3604711

Obenza, B. N., Salvahan, A., Rios, A. N., Solo, A., Alburo, R. A., & Gabila, R. J. (2024). University students' perception and use of ChatGPT: Generative artificial intelligence (AI) in higher education. International Journal of Human Computing Studies, 5(12), 5-18.

Oliver, R. L. (1980). A cognitive model of the antecedents and consequences of satisfaction decisions. *Journal of Marketing Research, 17*(4), 460–469. https://doi.org/10.1177/002224378001700405

Ouellette, J. A., & Wood, W. (1998). Habit and intention in everyday life: The multiple processes by which past behavior predicts future behavior. *Psychological Bulletin, 124*(1), 54–74. https://doi.org/10.1037/0033-2909.124.1.54

Plass, J. L., Homer, B. D., Mayer, R. E., & Kinzer, C. K. (2020). Theoretical foundations of game-based and playful learning. In J. L. Plass, R. E. Mayer, & B. D. Homer (Eds.), *Handbook of Game-based Learning* (pp. 3–24). MIT Press.

Schraw, G., & Olafson, L. (2008). Assessing teachers' epistemological and ontological worldviews. In M. S. Khine (Ed.), *Knowing, knowledge, and beliefs: Epistemological studies across diverse cultures* (pp. 25–44). Springer. https://doi.org/10.1007/978-1-4020-6596-5_2

Sobel, B. L. W. (2020). A new common law of web scraping. *Washington Law Review, 98*(3), 1219–1282.

Sousa, A. E., & Cardoso, P. (2025). Use of generative AI by higher education students. *Electronics, 14*(7), Article 1258. https://doi.org/10.3390/electronics14071258

Wang, T., Lund, B. D., Marengo, A., Pagano, A., Mannuru, N. R., Teel, Z. A., & Pange, J. (2023). Exploring the potential impact of artificial intelligence (AI) on international students in higher education: Generative AI, chatbots, analytics, and international student success. *Applied Sciences*, *13*(11), 6716.

Venkatesh, V., & Morris, M. G. (2000). Why don't men ever stop to ask for directions? Gender, social influence, and their role in technology acceptance and usage behavior. *MIS Quarterly, 24*(1), 115–139. https://doi.org/10.2307/3250981

Xiang, S., Deng, H., Wu, J., & Liu, J. (2024). Exploring the Integration of Artificial Intelligence in Research Processes of Graduate Students. In 2024 6th International Conference on Computer Science and Technologies in Education (CSTE) (pp. 110-113). *IEEE*.

Yin, C., Li, L., & Yu, L. (2024). Why do college students engage in in-class media multitasking behaviours? A social learning perspective. *British Journal of Educational Technology*, 55(3), 1105-1125.

Zhao, S., Luo, R., Sabina, C., & Pillon, K. (2023). The effect of information literacy training on graduate students' ability to use library resources. *College & Research Libraries*, *84*(1), 7.

## Appendix. Questions for Academic Information Source Preference

1. Which of the following best describes your academic major?
   a. Arts – fine art, music, dance, photography
   b. Humanities – philosophy, history, literature, languages
   c. Social Sciences – psychology, sociology, economics, education, library science
   d. Natural Sciences – biology, chemistry, mathematics, ecology, engineering
   e. Computing Sciences – computer science, information science, data science, AI
   f. Business – administration, hospitality, systems and decision sciences, marketing
2. What is your current academic standing?
   a. Undergraduate
   b. Master's Student
   c. Doctoral Student
3. What is your age?
   a. 18-25
   b. 26-30
   c. 31-35
   d. 36 and up
4. What is your gender?
   a. Female
   b. Male
   c. Non-Binary
   d. Other
5. Which best describes your student status?
   a. Domestic Student
   b. International Student
6. How many times have you engaged in the following activities in the past month?
   a. Used Google or similar search engine for university/course-related activities or assignments.
   b. Used an AI tool for university/course-related activities or assignments.
7. When you search for each of the following types of information, how often is the traditional Search Engine able to generate a satisfactory response? Note: A satisfactory response is one that fully addresses the information that you need and that you are "happy" with the information you received. (Never, not very often, somewhat often, very often, always):
   a. Finding job-related information
   b. Finding coursework-related information
   c. Learning more about a topic that you are studying
   d. Checking the weather
   e. Catching up on recent news
8. When you search for each of the following types of information, how often is the AI chatbot/ChatGPT able to generate a satisfactory response? Note: A satisfactory response is one that fully addresses the information that you need and that you are "happy" with the information you received. (Never, not very often, somewhat often, very often, always):

a. Finding job-related information
b. Finding coursework-related information
c. Learning more about a topic that you are studying
d. Checking the weather
e. Catching up on recent news